\documentclass[twocolumn,showpacs,amsmath,amssymb,prd,nofootinbib]{revtex4-1}
\usepackage{amsfonts}
\usepackage{amsmath}
\usepackage{amssymb}
\newcommand{\be}{\begin{equation}}\newcommand{\ee}{\end{equation}}
\newcommand{\bea}{\begin{eqnarray}}\newcommand{\eea}{\end{eqnarray}}
\newcommand{\bi}{\begin{enumerate}}
\newcommand{\ei}{\end{enumerate}}
\newcommand{\bref}[1]{(\ref{#1})}
\newcommand{\nn}{\nonumber}
\newcommand{\A}{\alpha}\newcommand{\B}{\beta} 
\newcommand{\D}{\delta}

\newcommand{\lam}{\lambda}\newcommand{\s}{\sigma}
          \newcommand{\w}{\omega}
          
\newcommand{\h}{\eta}

\def\6{\partial}\def\7{\widetilde}\def\8{\hat}\def\J{{\cal J}}
\def\pa{\partial}

\def\CL{{\cal L}}
\def\CF{{\cal F}}
\def\CK{{\cal K}}
\def\CJ{{\cal J}}
\def\={{\;=\;}}
\def\vs{\vskip 4mm}\def\+{{\;+\;}}

\def\swedge{{\wedge}}

\newcommand{\eq}[1]{eq.{\hskip 0.11cm{(\ref{#1})}}}

\newcommand\half{{\frac{1}{2}}}
\begin{document}
\preprint{}
\title{Generalized cosmological term from Maxwell symmetries }
\author{Jos\'e A. de Azc\'{a}rraga${}^1$}
\author{Kiyoshi Kamimura${}^2$}
\author{Jerzy Lukierski${}^3$}
\affiliation{${}^1$ Dept. of Theoretical Physics,
Univ. of Valencia and IFIC (CSIC-UVEG), 46100-Burjassot (Valencia), Spain}
\affiliation{${}^2$ Department of Physics, Toho University Funabashi
274-8510,  Japan }
\affiliation{${}^3$Institute of Theoretical
Physics, Wroclaw University, pl. Maxa Borna 9, 50-204 Wroclaw, Poland}

\begin{abstract}
By gauging the Maxwell spacetime algebra the standard geometric framework
of Einstein gravity with cosmological constant term is extended  by adding
six fourvector fields $A_\mu^{ab}(x)$ associated with the six abelian
tensorial charges in the Maxwell algebra. In the simplest Maxwell extension
of Einstein gravity this leads to a generalized cosmological term that
includes a contribution from these vector fields. We also consider
going beyond the basic gravitational model by means of bilinear actions for
the new Abelian gauge fields. Finally, an analogy with the supersymmetric
generalization of gravity is indicated. In an Appendix, we propose an
equivalent description of the model in terms of a shift of the standard
spin connection by the $A_\mu^{ab}(x)$ fields.
\end{abstract}

\maketitle


\section{Introduction}
It is known (see e.g. \cite{ajl1,ajl2}) that dark energy may be described
by adding the cosmological constant term to the standard Einstein-Hilbert
action. In a geometric framework leading to gravity, a cosmological term
appears when the de Sitter spacetime algebra is gauged. This algebra
contains (see e.g. \cite{ajl3}) noncommutative fourmomentum generators $P_{a}$,
$[P_a , P_b ] =  \frac{1}{R^2} \, M_{ab}\,$, where $M_{ab}$ are the six
Lorentz generators, $R$ is the de-Sitter radius and the cosmological constant
is identified as $\lambda=\frac{1}{R^2}$, $[\lambda]=M^2$.

A similar noncommutative modification of the Poincar\'e abelian fourmomenta
commutators also appears in the $D=4$ sixteen-dimensional Maxwell algebra
\cite{ajl4,ajl5}. This is given by
\begin{equation}\label{eqajl2}
[P_a , P_b ] =  \Lambda \, Z_{ab}  \; \,,
\end{equation}
where the six generators  $Z_{ab}$ ($a=0,1,2,3$) commute among themselves as
well as with $P_a$ and behave as an antisymmetric second rank Lorentz
tensor. The remaining Maxwell algebra commutators are
\begin{eqnarray}\label{eqajl3}
[Z_{ab}, Z_{cd}]&=& 0 = [P_{a}, Z_{cd}]\,,
\cr
[M_{ab}, P_{c}]&=& -(\eta_{ca} P_b - \eta_{cb} P_a) = -\eta_{c[a}P_{b]}\,,
\cr
[M_{ab}, Z_{cd}]&=& - (\eta_{c[a} Z_{b]d}- \eta_{d[a}Z_{b]c}) \; ,
\end{eqnarray}
plus the standard Lorentz algebra commutators for $M_{ab}$. Thus, the Maxwell
algebra has the semidirect sum structure
$\mathcal{I}\:{\supset \!\!\!\!\!\! \raisebox{1.5pt} {\tiny +}}\;so(1,3)\,$,
where the ideal $\mathcal{I}=\langle P_a,Z_{ab}\rangle $ is itself a central
extension of the abelian translation algebra $\langle P_a\rangle $ by
$\langle Z_{ab}\rangle $. The constant $\Lambda$
is dimensionful, $[\Lambda]=M^2$, and is the central charge that characterizes
the extension. Clearly, $[M_{ab}]=M^0$, $[P_a]=M$ and $[Z_{ab}]=M^0$.

Our aim in this paper is to consider an alternative way of introducing
the cosmological term.  This will appear in a generalized form, with a
dependence on the additional gauge fields associated with the new
generators $Z_{ab}$. In this paper we shall limit ourselves to providing
the new geometric framework; its applications to realistic cosmological
models will not be addressed in this paper. We shall consider the local
gauging of Maxwell algebra (\ref{eqajl2},\ref{eqajl3}) to look for possible
extensions of standard gravity. Because the non-commutativity
of the fourmomenta in de-Sitter gravity leads to the appearance
of a cosmological term, it is interesting to analyze the geometrical
consequences of the noncommutativity expressed by eq.~(\ref{eqajl2})
in a gauged Maxwell algebra approach to gravity. Further, since this
includes six gauge vector fields $A^{ab}_\mu$ associated with the
abelian $Z_{ab}$ generators, it is interesting to recall
(see {\it e.g.} \cite{ajl6, ajl7,ajl9}) that inflation can
also be driven by suitably coupled vector fields.

In this paper we introduce the geometric
framework obtained by gauging of the Maxwell group. Besides the
vierbein $e^a_\mu$ and the spin connection $\omega^{ab}_{\mu}$, our scheme
includes six vector fields $A^{ab}_{\mu}$ which introduce a
new set of curvatures. Besides the standard torsion $T^a$ corresponding
to the translational curvature, we have now two
curvature tensors, the standard Lorentz curvature tensor $R^{ab}_{\mu\nu}$ and
the new $F^{ab}_{\mu\nu}$ associated with
the six Abelian  gauge fields $A^{ab}_{\mu}$.
These two tensors will be the building blocks for constructing new
gravity actions. Our basic choice of the action will provide a
modification of the standard gravity, given by the Einstein action plus
a generalized cosmological term.
Our model will depend on three constants: the new central charge $\Lambda$
 in eq.~(\ref{eqajl2}), the conventional Einstein
gravitational constant
$\kappa$ ($[\kappa]=M^{-2}$) and the cosmological constant $\lambda$
($[\lambda]=M^{2}$) accompanying the standard cosmological term.

Additional gauge fields that describe the non-Riemannian
part of a connection have been considered in analysis
of metric affine gravity models (see \cite{He-Cr-Mi-Ne:95}, Sec. 3.11;
\cite{Bae-Bou-Heh:06}); the earliest example of a connection
modified by an abelian gauge field is the Weyl connection
\cite{Weyl:18}.
From these considerations it follows that one can use the one-forms
$A^{ab}=A^{ab}_\mu dx^\mu$ by formally extending the Riemannian connection
$\w^{ab}=\w^{ab}_\mu dx^\mu$ to non-Riemannian one with torsion
\be
\widetilde{\omega}^{ab}={\omega}^{ab}-\mu A^{ab}. \label{til-om}
\ee
We shall show further that the dimensionless parameter $\mu$ occurring in
\bref{til-om} is, in fact, equal to $\frac{\lambda}{\Lambda}$.
The antisymmetry $A_\mu^{ab}=-A_\mu^{ba}$ tells us that we are dealing with
an Einstein-Cartan geometry with non-metricity tensor equal to zero because
$\widetilde{\omega}^{(ab)}=0$ (a symmetric part of $\widetilde{\omega}^{ab}$
would define the non-metricity tensor \cite{Bae-Bou-Heh:06}).
As a result, the gauging of the Maxwell group may also be considered as the
specific extension to a non-Riemannian framework determined by the structure
of the Maxwell algebra.

The plan of the paper is the following. In Sect.2 we provide the differential
and geometric aspects of the gauging of Maxwell algebra. In Sect.3 we study
the Einstein action supplemented with the new generalized cosmological term,
which appears naturally in the present framework as a modification of the
standard four-volume form. We shall consider  further the field equations and
calculate the torsion generated by fields $A^{ab}_\mu$ as power series in
the parameter $\A=\frac{\mu^2}{\lambda}$. In order to have $A^{ab}_\mu$ as
dynamical fields we add an additional piece to the action for the new
Abelian gauge fields, as briefly discussed in Sect. 4.  To conclude, we
shall outline in Sect.5 some link between the structure of the Maxwell
generalization of gravity and the superextension of  gravity; we shall also
comment on the Maxwell extension of supergravity. The dynamics of Maxwell
gravity in terms of vierbein and the shifted spin connection
 $\7\w^{ab}$ in \bref{til-om} is given in Appendix A.

\section{ Gauging the Maxwell algebra.}

Let us introduce the set of Maxwell algebra-valued Maurer-Cartan forms
\begin{equation}
\label{eqajl4}
h= h^A X_A = e^a P_a + \frac{1}{2}\omega^{ab} M_{ab} +
\frac{1}{2}A^{ab} \, Z_{ab} \; ,
\end{equation}
where $a,b=0,1,2,3$ are tangent space indices raised and lowered with the
constant Minkowski metric $\eta_{ab}$. The associated gauge fields
$h^A_\mu (x)=\left( e^a_\mu(x), \omega^{ab}_{\mu}(x),
A^{ab}_{\mu}(x) \right)\,$ are defined by the $D=4$ spacetime one-form fields
\begin{equation}
\label{eqajl6}
e^a = e^a_{\mu} dx^\mu\;, \quad
\omega^{ab}=\omega^{ab}_\mu dx^\mu \; ,\quad
A^{ab} = A^{ab}_\mu dx^\mu \; ,
\end{equation}
where $(e^a_\mu , \omega^{ab}_{\mu})$ are the vierbein and the spin
connection and the $A^{ab}_\mu$ are the new abelian gauge fields;
$[e^a]=M^{-1}$, $[\omega^{ab}]=M^0$ and,
since $Z_{ab}$ is dimensionless, $[A^{ab}]=M^0$.

 The generators $X_A$=$(P_a , M_{ab}, Z_{ab})$ satisfy the Maxwell algebra
commutation relations, $[X_A , X_B]=f_{AB}{}^{C} X_C\,$. The generic
curvature two-forms of the associated gauge fields are given by
\bea
\label{eqajl7}
\mathcal{R}&=&dh+ h\swedge h = dh+\frac{1}{2}[h,h]\equiv \mathcal{R}^{A}X_A.
\eea
Denoting the components of $\mathcal{R}$ by
$\mathcal{R}^{A}=(T^a, R^{ab}, F^{ab})$, eqs.~(\ref{eqajl7})
and (\ref{eqajl2},\ref{eqajl3}) give
\begin{eqnarray}
T^a &=& d e^a+ \omega^a{}_c \swedge e^c \equiv (De)^a \,,
\label{eqajl13a}
\\
R^{ab} &=& d \omega^{ab} + \omega^a{}_{c} \swedge \omega^{cb} \equiv
(D\omega)^{ab}=-R^{ba} \,,
\label{eqajl13b}
\\
F^{ab} &=& d A^{ab} + \omega^{[a}_{\ \  c} \swedge A^{c|b]}
+\Lambda \ e^a \swedge e^b  \nn\\
&\equiv&  (DA)^{ab} + \Lambda  e^a
\swedge e^b =-F^{ba}  \,,
\label{eqajl13c}
\end{eqnarray}
where $D$ is the covariant derivative with respect to $\omega^{ab}$.
Eqs.~(\ref{eqajl13a},\ref{eqajl13b}) are the standard
torsion and curvature; eq.~(\ref{eqajl13c})  gives
the curvature $(DA)^{ab}$ of the abelian gauge fields $A^{ab}$
plus the vierbein two-form $\Lambda\, e^{a}\swedge e^{b}$.

Subsequently we obtain
\begin{eqnarray}
(DT)^a:&=& dT^a+ \omega ^a{}_c\swedge T^c = R^{ac} \swedge e_c \,,
\label{eqajl15a}
\\
(DR)^{ab}&=&(dR+\omega\swedge R-R\swedge\omega)^{ab} = 0 \,,
\label{eqajl15b}
\\
(DF)^{ab} &=& R^{[a|c} \swedge A_c^{\ b]} + \Lambda\ T^{[a} \swedge e^{b]} \;.
\label{eqajl15c}
\end{eqnarray}
Under a local gauge transformation with Maxwell algebra-valued
parameter $\zeta(x)$,
\begin{equation}
\label{eqajl9}
\zeta(x) = \zeta^{A}(x) \, X_{A}
=  \xi^a(x) P_a + \frac{1}{2}\lambda^{ab}(x) M_{ab}
+ \frac{1}{2} \rho^{ab}(x) Z_{ab}\ ,
\end{equation}
$h$ in eq.~(\ref{eqajl4}) transforms as
\begin{equation}
\label{eqajl11}
\delta_\zeta h^{\ A}
= d \zeta^A
+ f_{BC}^{\quad A} \, h^{\ B} \, \zeta^C  \equiv ({\cal D} \zeta)^A\; .
\end{equation}
Similarly, the curvatures in eq.~(\ref{eqajl7}) transform by
\begin{equation}
\label{eqajl12}
\delta_\zeta \mathcal{R^A} = f_{BC}{}^A \mathcal{R}^B\zeta^C \; ,
\end{equation}
which leads to
\begin{eqnarray}
\label{eqajl12b0}
\delta_\zeta e^a&=&(D\xi)^a+ e^c{\lambda_c}^a\ ,
\quad \delta_\zeta\omega^{ab}=(D\lambda)^{ab}\ ,
\\
\delta_\zeta A^{ab}&=&  (D\rho)^{ab}+A^{[a}{}_c\lambda^{c|b]}
+ \Lambda\ e^{[a}\xi^{b]}.
\end{eqnarray}
and
\begin{eqnarray}
\label{eqajl12b}
\delta_\zeta T^a&=&R^a{}_c\xi^c+ T^c{\lambda_c}^a\ ,
\quad \delta_\zeta R^{ab}={R^{[a}}_c\lambda^{c|b]}\ ,
\\
\delta_\zeta F^{ab}&=&  F^{[a}{}_c\lambda^{c|b]}
+R^{[a}{}_c\rho^{c|b]}+ \Lambda\ T^{[a}\xi^{b]}.
\end{eqnarray}
Thus, the two-forms $T^a,R^{ab}$ and $F^{ab}$ behave under local Lorentz
transformations $\lambda^{ab}(x)$ in a tensorial manner.

It follows from the above that dimensionless four-form
lagrangians invariant under diffeomorphism and the
local Lorentz transformations of the Einstein-Cartan theory
may be constructed as bilinears in $R^{ab}$ and $F^{ab}\,$,
\begin{eqnarray}
\mathcal{L}_1 &=& \half  \varepsilon_{abcd} R^{ab} \swedge R^{cd}
,\label{eqajl16a} \\
\mathcal{L}_2 &=& \varepsilon_{abcd} R^{ab} \swedge F^{cd}
,\qquad
\mathcal{L}_3 = \half \varepsilon_{abcd} F^{ab} \swedge F^{cd} \, .
\label{eqajl16c}
\end{eqnarray}
Further, we can consider as well
\begin{eqnarray}
\mathcal{L}_4 &=& \half  R^{ab} \swedge R_{ab}\,,
\label{eqajl17a}
\\
\mathcal{L}_5 &=&  R^{ab} \swedge F_{ab},\quad
\mathcal{L}_6 = \half F^{ab} \swedge F_{ab}\;.
\label{eqajl17c}
\end{eqnarray}
The terms (\ref{eqajl16a}) and (\ref{eqajl17a}) are known in a
standard gravity framework. The topological density
$\mathcal{L}_1$ produces a surface term which, in fact,
is proportional to the Euler characteristic. The term $\mathcal{L}_4$
is also topological and corresponds to the Chern-Pontrjagin class.
Our basic model will be constructed out of the lagrangian forms in
(\ref{eqajl16c}). \\

\section{ Einstein action with generalized cosmological term.}

Let us recall first that the Einstein-Hilbert action is
\begin{equation}
\label{eqajl18}
\mathcal{L}_E = -\frac{1}{2\kappa}\varepsilon_{abcd} \, R^{ab} \swedge e^c
\swedge e^d \; ,
\end{equation}
where $\kappa$ is the Einstein gravitational constant, $[\kappa]=M^{-2}$.
Then, it is seen that $\CL_2$ in \bref{eqajl16c} is
\begin{equation}
\label{eqajl19}
-\frac{1}{2\kappa \Lambda}\mathcal{L}_2
= -\frac{1}{2\kappa \Lambda}\varepsilon_{abcd} R^{ab} \swedge (DA)^{cd} +
\mathcal{L}_E \,.
\end{equation}
Now, using the Bianchi identity (\ref{eqajl15b})
the first term in the $r.h.s.$ of eq.~(\ref{eqajl19}) is
a surface term in the action:
\begin{eqnarray}
\label{eqajl20}
d(\varepsilon_{abcd} R^{ab} \swedge A^{cd}) &=&
\varepsilon_{abcd} R^{ab} \swedge (DA)^{cd} \; .
\end{eqnarray}
As a result, $\frac{-1}{2\kappa \Lambda}\mathcal{L}_2$ is the
Einstein-Hilbert lagrangian up to a surface term.

Let us now consider the $\CL_3$ in \bref{eqajl16c}  which is the announced
Maxwell extension of the cosmological term.
The standard cosmological term is given by the four-form
\begin{equation}
\label{eqajl21}
\mathcal{L}_{\rm {cosm}}
= \frac{\lambda}{4\kappa} \varepsilon_{abcd} \,
e^a \swedge e^b \swedge e^c \swedge e^d \;.
\end{equation}
If we observe that the curvature $F^{ab}$ is given by (\ref{eqajl13c}),
we see that $\mathcal{L}_3$ in eq.~(\ref{eqajl16c}) includes the standard
cosmological term plus two additional pieces depending on $A^{ab}$,
\begin{eqnarray}
\label{eqajl22}
&&\widetilde{\mathcal{L}}_{\rm {cosm}}
= \frac{\lambda}{2\kappa\Lambda^2} \, \mathcal{L}_3
 \cr
 &=&
 \frac{\lambda}{4\kappa\Lambda^2}  \varepsilon_{abcd}
 ((DA)^{ab} + \Lambda e^a \swedge e^b)\swedge
 ((DA)^{cd} + \Lambda e^c \swedge e^d)
 \cr
&=&
\mathcal{L}_{\rm {cosm}} +
\frac{\lambda}{4\kappa\Lambda^2} \varepsilon_{abcd} \, (DA)^{ab} \swedge
(DA)^{cd}
 \nn\\&&+ \frac{\lambda}{2\kappa\Lambda}
\varepsilon_{abcd} \, (DA)^{ab} \swedge e^c \swedge e^d\,.
\end{eqnarray}
Using eqs.~(\ref{eqajl16c}) and $\mu\equiv\frac{\lambda}{\Lambda}\ $, we
propose the following lagrangian four-form for Maxwell gravity
\begin{eqnarray}
\label{mod}
\mathcal {L}&=&\frac{\mu}{2\kappa\lambda}(-\mathcal{L}_2+\mu \mathcal{L}_3)
\nn\\&=&
\mathcal{L}_E + \mathcal{L}_{\mathrm{cosm}}
+\frac{\mu}{2\kappa}\varepsilon_{abcd}(DA)^{ab}\swedge e^c\swedge e^d
\nn\\&&
+\frac{\mu^2}{4\kappa\lambda}\varepsilon_{abcd}(DA)^{ab}\swedge (DA)^{cd}.
\end{eqnarray}

Let us compute the field equations.
The variation of the Lagrangian \bref{mod}  with respect to $\w^{ab}$ gives
\bea
\D_\w\CL&=&\D\w^{ab}\swedge [L]_{\omega^{ab}}
\nn\\&=& d[-\frac{1}{2\kappa}\varepsilon_{abcd} \, \D\w^{ab} \swedge e^c
\swedge e^d]-\frac{1}{\kappa}\varepsilon_{abcd} \,\D\w^{ab} \swedge (De)^c
\swedge e^d
\nn\\&+&\frac{\mu}{\kappa}\varepsilon_{abcd}{\D\w^a}_e\swedge A^{eb}\swedge
e^c\swedge e^d
\nn\\&&
+\frac{\mu^2}{\kappa\lambda}\varepsilon_{abcd}{\D\w^a}_e\swedge A^{eb}\swedge
 (DA)^{cd}
\nn\\&=& \D\w^{ab}[-\frac{1}{\kappa}\varepsilon_{abcd} \, \swedge [
(De)^c \swedge e^d-\frac{\mu^2}{\lambda}{A^c}_{e}\swedge (
 (DA)^{ed}
\nn\\&&+\frac{\lambda}{\mu}e^c\swedge e^d)]].
\end{eqnarray}
We then obtain
\bea
[L]_{\omega^{ab}}=-\frac{1}{\kappa}\varepsilon_{abcd} \,
 [(De)^c \swedge e^d-\frac{\mu^2}{\lambda}{A^c}_{e}\swedge
 F^{ed}]=0. \label{harto3}
\label{tormu00}\end{eqnarray}
 The equation \bref{tormu00} expressed in terms of the standard torsion
$T^a=(De)^a$ is the
following
\be
\label{tormu2}
T^{[a}\swedge e^{b]} +\frac{\mu^2}{\lambda} F^{[a}{}_c \swedge A^{c|b]} = 0.\;
\ee
It will be further used as the algebraic equation determining the spin
connection as functions of vierbein and new gauge fields; $\w^{ab}_\mu(e,A)$.

The variation of  \bref{mod}  with respect to  $e^a$ gives
\bea
\D_e\CL&=&\D e^a\swedge [L]_{e^a}
\nn\\&=& -\frac{1}{\kappa}\varepsilon_{abcd} \, R^{ab} \swedge e^c \swedge
\D e^d
+ \frac{\lambda}{\kappa} \varepsilon_{abcd} \,
e^a \swedge e^b \swedge e^c \swedge \D e^d
\nn\\&&+\frac{\mu}{\kappa}\varepsilon_{abcd}(DA)^{ab}\swedge e^c\swedge \D e^d
\nn\\&=& -\frac{1}{\kappa}\D e^a \varepsilon_{abcd} \,\swedge [ R^{bc}
\swedge e^d -{\lambda} \,e^b \swedge e^c \swedge e^d
-{\mu} (DA)^{bc}\swedge e^d]\nn\\
\label{vieerq200}\end{eqnarray}
so that, using \bref{eqajl13c},
\be
[L]_{e^a}= -\frac{1}{\kappa} \varepsilon_{abcd} \,[ R^{bc} \swedge e^d
-{\mu} F^{bc}\swedge e^d]=0.\label{vieerq20}
\ee
The curvature satisfies the field equation
\begin{eqnarray}\label{vieerq2}
\varepsilon_{abcd}\,e^b\wedge\left( R^{cd}-\lambda e^c\swedge e^d
 - \mu (DA)^{cd}\right)=0 .
\eea

The variation of  \bref{mod}  with respect to $A^{ab}$ gives
\bea
\D_A\CL&=&\D A^{ab}\swedge [L]_{A^{ab}}\nn\\
&=&d[\frac{\mu}{2\kappa}\varepsilon_{abcd}\D A^{ab}\swedge e^c\swedge e^d +
\frac{\mu^2}{2\kappa\lambda}\varepsilon_{abcd}\D A^{ab}\swedge (DA)^{cd}]
\nn\\&+&
\frac{\mu}{\kappa}\varepsilon_{abcd}\D A^{ab}\swedge (De)^c\swedge e^d
+\frac{\mu^2}{2\kappa\lambda}\varepsilon_{abcd}\D A^{ab}\swedge (DDA)^{cd}
,\nn\\ \eea
from which it follows that
\be
[L]_{A^{ab}}=\frac{\mu}{\kappa}\varepsilon_{abcd}[ (De)^c\swedge e^d
+\frac{\mu}{\lambda}{R^{c}}_e\swedge A^{ed}]=0. \label{harto3A}
\ee
Eq. \bref{harto3A} can be written alternatively using the torsion as
\be
T^{[a}\swedge e^{b]}+\frac{\mu}{\lambda}{R^{[a}}_e\swedge A^{e|b]}=0.
\label{tormu23}\ee

A special solution of \eq{vieerq2} is given by
\be\label{RFrel}
R^{ab}=\mu\, F^{ab}=\mu (DA)^{ab}+\lambda e^a\swedge e^b.
\ee
If \eq{RFrel} holds, after using the Bianchi identities \bref{eqajl15b} and
\bref{eqajl15c} one obtains \eq{tormu23}, which can be rewritten as
\be\label{DRF}
(DF)^{ab}=0.
\ee
Further, if we insert \eq{RFrel} in \eq{tormu23} we get \eq{tormu2}.
 We see therefore that the set of equations of motion \bref{tormu2},
\bref{vieerq2} and  \bref{tormu23} are satisfied if the Lorentz and
gauge connections are related by \bref{RFrel}.

Let us now solve \eq{harto3} or  \eq{tormu2} by expressing $\w^{ab}$
in terms of the vierbein and $A^{ab}$.  First we note that eqs. \bref{harto3}
 are six three-form equations
\bea
\varepsilon_{abcd} [
(De)^c \swedge e^d-\frac{\mu^2}{\lambda}{A^c}_{e}\swedge (
 (DA)^{ed}+\frac{\lambda}{\mu}e^c\swedge e^d)]=0, \label{harto4}
\nn\\ \end{eqnarray}
depending linearly on the 24 unknowns $\w_\mu^{ab}$. Since the number of
equations and unknowns match, in principle \eq{harto4}
 can be solved algebraically.
We recall that in the standard gravity $(\mu=0)$ the equation
\be
\varepsilon_{abcd} (De)^c \swedge e^d=0,\quad\to\quad
T^c= (De)^c=de^c+\w^{cd}\swedge e_d=0,
\ee
is solved assuming regularity of ${e_\mu}^a$ as
\bea
\w_{ab}&=&\w^{(0)}_{ab}=\frac12(W_{bc,a}+W_{ca,b}-W_{ab,c}){e}^c,
\label{tfree}\qquad \nn\\ &&
W_{ab,c}\equiv{e_a}^\rho {e_b}^\s\pa_{[\rho}{e_{\s]c}}.
\label{w0ab}\eea
Eq. \bref{harto4} is simpler if we use the shifted connection
$\7\w^{ab}=\w^{ab}-\mu A^{ab}$ (see \eq{til-om})
\begin{eqnarray}\label{harto51}
\varepsilon_{abcd}[{\7\w^{ae}}\swedge (e_e\swedge  e^b&+&\frac{\mu^2}
{\lambda}  A_{ef}\swedge A^{fb})
\nn\\
+de^a \swedge e^b
&+&\frac{\mu^2}{\lambda} dA^{ae}\swedge {A_{e}}{}^{b}]=0,
\eea or, equivalently,
\bea
\frac12\varepsilon_{abcd}\left(d{\CK}^{ab}+{\7\w^{[ae}}\swedge {{\CK}_e}^{b]}
\right)=0,
\label{weqat}\eea
where
\be
{\CK}^{ab}=e^a \swedge e^b+\frac{\mu^2}{\lambda}  {A^a}_{f}\swedge A^{fb}.\ee
\vs
We may now find a perturbative solution of \eq{harto51} for
$\w^{ab}$. First, we write $\7\w_{ab}=\w^{(0)}_{ab}+\A\w^{(1)}_{ab}+\A^2
\w^{(2)}_{ab}+...$ or, equivalently,
\be
\w_{ab}=\mu A_{ab}+\w^{(0)}_{ab}+\A\w^{(1)}_{ab}+\A^2\w^{(2)}_{ab}+...
\label{wser}\ee
where $\A=\frac{\mu^2}{\lambda}$ and $\w^{(0)}_{ab}$ is given in \eq{tfree}.
Inserting \bref{wser} in \eq{harto4} we find
\bea
&&\varepsilon_{abcd} [
(de^c+\w^{ce}\swedge e_e)\swedge e^d+\A( dA^{ce}+{\w^{cf}}\swedge {A_f}^e)
\swedge {A_e}^d
\nn\\&&-{\mu}{A^c}_{e}\swedge e^e \swedge e^d]
\nn\\&=&
\;\; \varepsilon_{abcd} [
(\w^{(0)}\swedge e)^{c}\swedge e^d+de^c\swedge e^{d}]
\nn\\&+&
\A\, \varepsilon_{abcd} [
(\w^{(1)}\swedge e)^{c}\swedge e^d+(dA\swedge A)^{cd}+(\w^{(0)})^{ce}\swedge
{(A\swedge A)_e}^d]
\nn\\&+&
\sum_{i=2}^\infty \A^i\, \varepsilon_{abcd} [
(\w^{(i)}\swedge e)^{c}\swedge e^d+(\w^{(i-1)})^{ce}\swedge {(A\swedge A)_e}^d]
=0. \nn\\ \label{harto52}
\end{eqnarray}
Requiring that the terms for different powers of $\A$ should vanish separately
we can determine recursively $\w^{(n)}$.
This defines the standard torsion as follows
\bea
T^a&=&de^a+\w^{ab} \swedge e_b
=(\mu A+\A\w^{(1)}+\A^2\w^{(2)}+...)^{ab}\swedge  e_b.
\nn\\ \eea
The $\A^0$ term in \eq{harto52} vanishes if we choose $\w^{(0)}_{ab}$ as given
by \eq{w0ab}; the other terms $\w^{(j)}_{ab}(j>0)$ follow recursively
and depend on the gauge fields $A_\mu^{ab}$ and their derivatives
(see Appendix B).

By solving \eq{tormu2}, we can eliminate the spin connection $\w^{ab}_\mu$
and move to a second order formalism, with independent variables
$e^{a}_\mu$ and $A^{ab}_\mu$. At the next step the differential
equations~\bref{vieerq2} and  \bref{tormu23} are solved for $e^{a}_\mu$ and
$A^{ab}_\mu$.
It is worth noting that \eq{vieerq2} adopts the form of a generalized
Einstein equation for the shifted curvature
\be\label{JeqRF}
J^{ab}\equiv R^{ab}-\mu F^{ab}.
\ee
After expressing $J^{ab}$ in local coordinates
\eq{vieerq2} takes the form
\begin{equation}
{\J^\mu}_\nu-\frac12{\delta^\mu}_\nu\,\J=({R^\mu}_\nu-\mu{\CF^\mu}_\nu)-
\frac12{\delta^\mu}_\nu(R-\mu\CF)=0
\label{EinsteinJ0}\end{equation}
or,  equivalently,
\begin{equation}
{G^\mu}_\nu={R^\mu}_\nu-\frac12{\delta^\mu}_\nu\,R=\mu\,{T^\mu}_\nu,
 \qquad {T^\mu}_\nu\equiv {\CF^\mu}_\nu-\frac12{\delta^\mu}_\nu\,\CF.
\label{EinsteinJR}\end{equation}
Here
\bea
{e_a}^\mu {e_b}^\nu J^{ab}&=&\frac12{\J^{\mu\nu}}_{\rho\sigma}dx^\rho \wedge
dx^\sigma
,\nn\\&& \quad {\J^{\mu}}_{\rho}\equiv {\J^{\mu\nu}}_{\rho\nu}
,\quad \J\equiv {\J^{\mu}}_{\mu},
\\ \label{Defrieman}
{e_a}^\mu {e_b}^\nu R^{ab}&=&\frac12{R^{\mu\nu}}_{\rho\sigma}dx^\rho
\wedge dx^\sigma,\nn\\&& \quad
{R^{\mu}}_{\rho}\equiv{R^{\mu\nu}}_{\rho\nu},\quad R\equiv{R^{\mu}}_{\mu},
\eea
\bea
{e_a}^\mu {e_b}^\nu F^{ab}&=&
\frac12{\CF^{\mu\nu}}_{\rho\sigma}dx^\rho \wedge dx^\sigma\,,
\label{EinsteinJR54}\\
{\CF^{\mu}}_{\rho}\equiv{\CF^{\mu\nu}}_{\rho\nu}&=&
{e_a}^\mu {e_b}^\nu(D_{[\rho}A_{\nu]})^{ab}+3\Lambda {\delta^\mu}_\rho,
\label{EinsteinJR55}\\
 \CF\equiv {\CF^{\mu}}_{\mu}&=&{e_a}^\mu {e_b}^\nu(D_{[\mu}A_{\nu]})^{ab}+12\Lambda, \label{EinsteinJR56}
\eea
where ${R^{\mu\nu}}{}_{\rho\s}, {R^{\mu}}{}_{\rho}, R$ are the Riemann,
Ricci and scalar curvatures and $D_{\mu}$ is the covariant derivative with
respect to $\omega^{ab}_\mu$, which are now given as functions of $e^a_\mu$
and $A^{ab}_\mu$.
Using \eq{EinsteinJR55}-\bref{EinsteinJR56}  equation \bref{EinsteinJR} may
be written as the Einstein equation in de-Sitter space with cosmological
constant $\lambda=\mu\Lambda$,
\bea
&&{R^\mu}_\nu-\frac{R}2\,{\delta^\mu}_\nu\,-3\,\lambda\,{\delta^\mu}_\nu
\nn\\&=&
\mu\,\left({e_a}^\mu {e_b}^\s(D_{[\nu}A_{\s]})^{ab}-{\delta^\mu}_\nu
{e_a}^\rho {e_b}^\s(D_{\rho}A_{\s})^{ab}\right)
\label{EinsteinJR57}\eea
with the source linear in new gauge fields.

We conclude this section by noting that in
Appendix A we show that the action \bref{mod} and its equations of motion
may be equivalently described with the use of shifted spin connection $\7\w^{ab}$ and
curvature $J^{ab}$.

\section{Dynamical terms for the new gauge fields}

The remaining equation \bref{harto3A}, obtained by varying the action
\bref{mod} with respect to the fields $A_\mu^{ab}$, does not depend
explicitly on the derivatives of $A^{ab}_\mu$. In order to have dynamical
gauge fields $A_\mu^{ab}$, terms bilinear in their derivatives are
needed. In the collection of diffeomorphism invariant geometrical
actions \bref{eqajl16a}-\bref{eqajl17c} only the term $\CL_6$ could be a
candidate, but due to formula \bref{eqajl13c} its non-topological part is
only linear in $A_\mu^{ab}$. Thus, to get the free action for new gauge fields
a Maxwell-like term $\7\CL_6=-\frac{\B}2\,F\swedge*F$ would have to be added,
 however it is less geometric since the Hodge star operator involves
the metric $g_{\mu\nu}=\h_{ab}{e_\mu}^a {e_\nu}^b$. It takes the form
\be
\7\CL_6=-\,\B\frac{\sqrt{-g}}4g^{\mu\nu}g^{\rho\s}F_{\mu\rho}^{ab}
F_{\nu\s ab}\,d^4 x\,,
\label{L66}\ee
where $g=\det(g_{\mu\nu})$.

The field equations following from the addition of \bref{mod} and
\bref{L66} look as follows
\bea
\D \w^{ab}&:&\;-\frac{1}{\kappa}\varepsilon_{abcd} \,
 [(De)^c \swedge e^d-\frac{\mu^2}{\lambda}{A^c}_{e}\swedge F^{ed}]
\nn\\&&\hskip 25mm -\B\,{A}_{c[a}\swedge {(*F)_{b]}}^c=0,
\label{exteoms2}\\
\D e^a&:&\;-\frac{1}{\kappa} \varepsilon_{abcd} \,[ R^{bc} \swedge e^d
-{\mu} F^{bc}\swedge e^d]\nn\\&&\hskip 5mm -2\B\,\Lambda\,(*F)_{ab}\swedge
e^b-\B\,{T_{Fa}}^b\,*e_b=0,
\label{exteoms1}\\
\D A^{ab}&:&\;\frac{\mu}{\kappa}\varepsilon_{abcd}[ (De)^c\swedge e^d
+\frac{\mu}{\lambda}{R^{c}}_e\swedge A^{ed}]\nn\\&&\hskip 30mm
-\B\,(D*F)_{ab}
=0, \label{exteoms3}\eea
where ${T_{Fa}}^b$ is
\be
{T_{Fa}}^b={e_{\mu a}}{e_\nu}^b\,\left(\frac{g^{\mu\nu}}{4}(F^{\rho\s}F_{\rho\s})-
\frac12 F^{(\mu\rho}{F^{\nu)}}_\rho\right).
\ee
Equation \bref{exteoms2} modifies the torsion relation (\eq{tormu2})
 and changes
the expression for the spin connection $\w_\mu^{ab}$ in terms of $e^a_\mu$
and $A^{ab}_\mu$ (see Appendix B for the $\B=0$ case).
Equation \bref{exteoms1} modifies the energy momentum tensor in
\eq{EinsteinJR}.
Finally,
\eq{exteoms3} produces a dynamical equation for  $A^{ab}_\mu$. If we use
Bianchi identity,   \bref{eqajl15b} and \bref{eqajl15c} ,
\eq{DRF} is replaced by the following one
\be
(DF)^{ab}=-\frac{\B\kappa\Lambda}{2\mu} \varepsilon^{abcd}(D*F)_{cd}.
\ee

\section{ Final remarks.}

It is often thought that the cosmological constant problem may require
an alternative approach to gravity. Here we have presented a new
geometric framework, based on the $D=4$ Maxwell algebra
\cite{foot1}, which involves six new gauge fields associated
with their abelian generators, and described its simplest application:
a generalization of the cosmological term.

There are some possible extensions of this work, as

a) Using the analogy between the structure of the Maxwell
and supersymmetry algebras,
\be
\nonumber
\langle P_a,Z_{ab}\rangle {\supset \!\!\!\!\!\! \raisebox{1.5pt} {\tiny +}}\
so(1,3) \quad ,
\quad
\langle Q_\alpha,P_\mu\rangle {\supset \!\!\!\!\!\! \raisebox{1.5pt}
{\tiny +}}\ so(1,3)\; ,
\ee
we can obtain the bosonic Maxwell counterpart of the superspace formulation
of supergravity by enlarging spacetime with
the Maxwell group variables associated with the $Z_{ab}$ generators.

b) Recently, the simplest Maxwell superalgebra was introduced
in \cite{BGKL:09}. This algebra could be gauged following the
approach presented in this paper to provide
an extension of the standard $D=4$ supergravity
framework. Besides the fields $A_\mu^{ab}(x)$, such an
approach would include two gravitino fields: the standard gravitino
and an additional one, required by the two Weyl charges
in the Maxwell superalgebra \cite{BGKL:09}.

c) An important step in extending the model presented here would consist
in adding covariantly coupled matter fields as sources,
which would appear as local currents on the
$r.h.s.$ of the equations for the Maxwell gravity gauge fields.
As it is known, the equation for the spacetime curvature has the
energy-momentum tensor as its source, and the torsion is coupled to the
local spin density. In order to  introduce the new local currents
describing the sources of the additional gauge fields $A_\mu^{ab}$
we should couple them to matter invariant under the Maxwell symmetry.
The new
local currents would define the local densities providing, after space
integration, the conserved tensorial central charges $Z_{ab}$.
\\

{\it Acknowledgements.} The authors wish to thank  Joaquim Gomis
for his interest in the subject and valuable comments at the
initial stages of this paper, as well as support from
the research grants FIS-2008-1980 from the Spanish
MICINN and NN202-331139 from the Polish Min. of Science and
Higher Education.

\appendix

\section{Maxwell gravity in terms of the shifted Riemannian connection $\7\w^{ab}$}

We may also add to the lagrangian \bref{mod} the topological density
in eq.~(\ref{eqajl16a}) as follows
\begin{equation}
\label{eqajl23}
\mathcal{L} = \frac{1}{2\kappa\lambda} ( \mathcal{L}_1 -\mu\ \mathcal{L}_2 +
\mu^2\, \mathcal{L}_3 )
\end{equation}
since $ \mathcal{L}_1$ is a surface term only the last two terms contribute
to the field equations.
Therefore, the $\mathcal{L}$ in eq.~(\ref{eqajl23}) may be expressed as
a quadratic expression in the $R(\omega)$ curvature shifted by bilinear
terms in the vierbein \cite{ajl3, Obu-He:96, Ed-Ha-Tro-Za:06} and
by the new gauge fields $A^{ab}$,
\begin{equation}
\label{eqajl24}
\mathcal{L}= \frac{1}{4\kappa\lambda} \, \varepsilon_{abcd}  J^{ab} \swedge
J^{cd}\;,
\end{equation}
where $J^{ab}$ is given in \eq{JeqRF}.
Denoting $(A^2)^{ab}={A^a}_c\swedge A^{cb}$, we get
\begin{eqnarray}
\label{eqajl25}
J^{ab} &=&R^{ab}(\omega) - \mu\, F^{ab}
=R^{ab}(\widetilde{\omega})-\lambda e^a \swedge e^b- \mu^2 \, (A^2)^{ab}
\cr &
\equiv& \widetilde{R}^{ab} - \lambda e^a\swedge e^b,
\end{eqnarray}
where  $\widetilde{\omega}^{ab}$ is given in  eq.~(\ref{til-om}) and
\begin{equation}
\label{eqajl26}
R^{ab}(\widetilde{\omega})
\equiv  d \widetilde{\omega}^{ab}
+ \widetilde{\omega}^{a}{}_c \swedge \widetilde{\omega}^{cb} \; , \;
\widetilde{R}^{ab} \equiv R^{ab}(\widetilde{\omega}) - \mu^2 (A^2)^{ab}\; .
\end{equation}
Note that it is $\widetilde{R}^{ab}$ rather than $R^{ab}(\widetilde{\omega})$
that is the `true' curvature of the shifted connection
$\widetilde{\omega}^{ab}$,
since $\widetilde{R}^{ab}$ does not contain (because the $Z_{ab}$ are abelian)
the $\mu^2 (A^2)^{ab}$ piece that is present in $R^{ab}(\widetilde{\omega})$.

The lagrangian $\mathcal{L}$ in (\ref{eqajl24}) may then be written in
the following two equivalent forms
\be
\mathcal{L}= \varepsilon_{abcd}(\frac{1}{4\kappa\lambda}\widetilde{R}^{ab}
\swedge\widetilde{R}^{cd}
- \frac{1}{2\kappa}\widetilde{R}^{ab}\swedge e^c\swedge e^d
+\frac{\lambda}{4\kappa}e^a\swedge e^b \swedge e^c \swedge e^d)
\ee
and
\begin{eqnarray}
\label{k30}
\mathcal{L} &=&
\frac{1}{4\kappa\lambda}\varepsilon_{abcd}R^{ab}(\widetilde{\omega})
\swedge R^{cd}(\widetilde{\omega})
-\frac{1}{2\kappa}\varepsilon_{abcd}\,e^a\swedge e^b\swedge
R(\widetilde{\omega})^{cd}\cr
 &+&\frac{\lambda}{4\kappa}\varepsilon_{abcd}\,e^a\swedge e^b\swedge e^c
\swedge e^d+\frac{\mu^4}{4\kappa\lambda} \varepsilon_{abcd}(A^2)^{ab}
\swedge(A^2)^{cd}\,\cr
 &-&\frac{\mu^2}{2\kappa\lambda}\varepsilon_{abcd}(R^{ab}(\widetilde{\omega})-
\lambda e^a\swedge e^b)(A^2)^{cd}.
\end{eqnarray}
The first term in (\ref{k30}) is an exact form and will be ignored.
The second piece of $\mathcal{L}$ is the Einstein-Hilbert action for
the shifted connection $\widetilde{\omega}$ and the third one is the
standard cosmological term. The fourth term
of $\mathcal{L}$ vanishes  due to the identity
\begin{equation}
\label{ideA}
\varepsilon_{abc[d} (A^2)^{ab}\swedge A^c{}_{e]}=0 \; ,
\end{equation}
that holds for any antisymmetric one-form $A^{ab}$.
Finally, the last term is the remaining
addition to the standard cosmological term. Thus, we can write
\begin{equation}
\label{finlag}
\mathcal{L}= \mathcal{L}_{EH}(\widetilde{\omega})
+\mathcal{L}_{\rm{cosm}}+\mathcal{L}_A \;,\qquad
\ee
\be
\mathcal{L}_A  =
-\frac{\mu^2}{2\kappa\lambda}\varepsilon_{abcd}(R^{ab}(\widetilde{\omega})
-\lambda e^a\swedge e^b)\swedge (A^2)^{cd} \;.
 \end{equation}

Let us now consider the field equations, obtained by varying
$I=\int\mathcal{L}\,$ with respect to $\widetilde\omega^{ab}, e^a$ and
$A^{ab}$.
\begin{eqnarray}
\label{harto}
\delta\widetilde{\omega}^{cd}&:&
\;\varepsilon_{abcd}\left((\widetilde{D}e)^a{\swedge} e^b
+\frac{\mu^2}{\lambda} (\widetilde{D}A)^{ae}\swedge {A_{e}}{}^{b}\right)=0,
\\
\label{vieerq}
\delta e^a &:&
\varepsilon_{abcd}\,e^b\wedge\left( R^{cd}(\widetilde{\omega})
-\lambda e^c\swedge e^d - \mu^2 (A^2)^{cd}\right)=0 ,
\nn\\ \\
\label{deltaA}
\delta A^{de}&:&
\varepsilon_{abc[d}(R^{ab}(\widetilde{\omega})
- \lambda e^a \swedge e^b)\swedge A^c{}_{e]} =0 \;.
\end{eqnarray}
Due to identity \bref{ideA}, equation \bref{deltaA} can be replaced by
\begin{equation} \label{deltaA2}
\varepsilon_{abc[d}(R^{ab}(\widetilde{\omega})
- \lambda e^a \swedge e^b-\mu^2(A^2)^{ab})\swedge A^c{}_{e]} =0 \;.
\end{equation}
The Bianchi identity for $R(\widetilde{\omega})^{ab}$,
$(\widetilde{D}R(\widetilde{\omega}))^{ab}=0$, shows
\begin{equation}
\label{4seis}
(\widetilde{D} J)^{ab}=
-\lambda (\widetilde{D} e)^{[a}\swedge e^{b]}
-\mu^2(\widetilde{D} A)^{[a}{}_c \swedge A^{c|b]}  \; .
\end{equation}
Using  it in eq.~\bref{harto}
the set of equations of motion becomes
\begin{eqnarray}
\delta \widetilde\omega^{ab}: \quad  && (\widetilde{D}J)^{ab}
= dJ^{ab}+\widetilde{\omega}^{[a|c}{J_c}^{b]}= 0\,,
\label{7cuatrob} \\
\delta e^a: \quad &&
\varepsilon_{abcd} e^b J^{cd} =0\,,
\label{7cuatroa}\\
\delta A^{de}: \quad && \varepsilon_{abc[d} J^{ab} A^c{}_{e]}= 0 \;.
 \label{7cuatroc}
\end{eqnarray}
They coincide with the equations of motion \bref{tormu2}, \bref{vieerq2} and
\bref{tormu23} respectively.

Writing the forms in local coordinates (see also \eq{Defrieman})
\be
{e_a}^\mu {e_b}^\nu J^{ab}=\frac12{\J^{\mu\nu}}_{\rho\sigma}dx^\rho \wedge
dx^\sigma
,\quad {e_a}^\mu {e_b}^\nu A^{ab}=\frac12{A^{\mu\nu}}_{\rho}dx^\rho,
\ee
after assuming the invertibility for the vierbein, we obtain
\bea
{\J^{\mu\nu}}_{\rho\s}&=&{R^{\mu\nu}}{}_{\rho\s}(\widetilde\omega)-\lam
{\delta^\mu}_{[\rho}{\delta^\nu}_{\s]}-\mu^2
{A^{\mu \lambda}}_{[\rho}{{A_{\lambda}}^\nu}_{\s]},
\label{Jmnrs}\\
{\J^{\mu}}_{\rho}&\equiv&{\J^{\mu\nu}}_{\rho\nu}\nn\\&=&
{R^{\mu}}{}_{\rho}(\widetilde\omega) -3\,\lam\,{\D^\mu}_{\rho}
+\mu^2\,({A^{\mu \lambda}}_{\rho}{A^\nu}_{\lambda\nu}-
{A^{\mu \lambda}}_{\nu}{A^\nu}_{\lambda\rho}),
\nn\\ \label{Jmn}\\
\J&=&{\J^{\mu}}_{\mu}\nn\\&=& R(\widetilde\omega)-12\,\lam
+\mu^2\,({A^{\mu \lambda}}_{\mu}{A^\nu}_{\lambda\nu}
-{A^{\mu \lambda}}_{\nu}{A^\nu}_{\lambda\mu}),
\nn\\ \label{J}\eea
where ${R^{\mu\nu}}{}_{\rho\s}(\widetilde\omega),
{R^{\mu}}{}_{\rho}(\widetilde\omega), R(\widetilde\omega)$ are the Riemann,
Ricci, and scalar tensors for the shifted spin connection $\widetilde\omega$.
By following the derivation of Einstein equation
from the  Einstein-Hilbert Lagrangian \bref{eqajl18},
we obtain  the generalized  Einstein equation \eq{EinsteinJ0},
with ${\CJ^\mu}_\nu$ and $\CJ$ expressed by the formulae
in \bref{Jmn} and  \bref{J}.

An obvious solution of eqs.~\bref{7cuatrob}-\bref{7cuatroc} is
$J^{ab}=0$ (see also \eq{RFrel}), which
in the formalism with shifted spin connection, specifies the curvature
through eq.~(\ref{eqajl25}) as
\begin{equation}
\label{cuatro1}
R(\widetilde{\omega})^{cd}=\lambda\, e^c\swedge e^d+\mu^2(A^2)^{cd}\;.
\end{equation}
In such a case the new gauge fields are arbitrary, not restricted by
 eq.~(\ref{7cuatroc}).  If, however, $J^{ab}\neq0$, the explicit solutions of
 the generalized Einstein equation \bref{EinsteinJ0} will
 then provide a restriction on the abelian gauge fields
 $A^{ab}_\mu$ since eq.~(\ref{7cuatroc}) will no longer be trivial.

We mention that to the Lagrangian \bref{eqajl23}
one can add new terms by using the lagrangian
densities (\ref{eqajl17a}-\ref{eqajl17c}) as follows
\be
\mathcal{L}'=
\frac{a}{2 \kappa\lambda}(\mathcal{L}_4 - \mu \mathcal{L}_5+ \mu^2
\mathcal{L}_6) \;,
\ee
where $ a $ is a dimensionless constant. The total lagrangian
becomes
\be
\mathcal{L}+\mathcal{L}' = \frac{1}{4\kappa\lambda}(\epsilon_{abcd}J^{ab}
\swedge J^{cd}
+ a  J_{ab}\swedge J^{ab}) \; ,
\label{addLagB}
\ee
which leads to eqs.~(\ref{7cuatrob}-\ref{7cuatroc}) but written now the
tensor $\widetilde{J}^{ab}=J^{ab}-\frac{ a }{4}\varepsilon^{ab}{}_{cd}
J^{cd}$. As mentioned in the main text,
the lagrangian \bref{addLagB} does not contain a `free' term for the
$A^{ab}$ fields; this may be achieved by adding a ($F\wedge * F$)-type
term, as in Sect.4, which is not among the densities
considered in eqs.~(\ref{eqajl16a}-\ref{eqajl17c}).\\

\section{Expression for the higher ${\w^{(j)ab}}$}

The explicit expression for the higher order terms are determined
recursively as follows.
We write \eq{harto52}, for $ j=0,1,2,...$, as
\bea
 \varepsilon_{abcd} {\w^{(j)}}^{ce}\swedge e_{e}\swedge e^{d}+
K^{(j)}_{ab}=0,
\eea
where
\bea
K^{(0)}_{ab}&=& \varepsilon_{abcd}\,de^c\swedge e^d,\quad
\nn \\ K^{(1)}_{ab}&=& \varepsilon_{abcd}(dA^{ce}+\w^{(0)[cf}\swedge{A_f}^{e]})\swedge
{A_e}^{d},\nn\\
K^{(i)}_{ab}&=& \varepsilon_{abcd}\,
\w^{(i-1)ce}\swedge {(A\swedge A)_e}^d
,\quad (i=2,3,...).
\eea
If we express the three-form $K^{(j)}_{ab}$ in terms of the three-forms
$*e_c$ as
\be
K^{(j)}_{ab}={K^{(j)}_{ab,c}}(*e^c), \qquad e^a\swedge e^b\swedge e^c\equiv
\varepsilon^{abcd}(*e_d),\ee
we find that $\w^{(j)}_{ab}$ is given by
\be
\w^{(j)}_{ab}=\frac12\left((K^{(j)}_{bc,a}+K^{(j)}_{ca,b}-K^{(j)}_{ab,c})e^c+
K^{(j),e}_{[ae}e_{b]}\right)=-\w^{(j)}_{ba}.
 \label{harto71}\ee
For $j=0$ this recovers \bref{w0ab}. For $j>0$, the $\w^{(j)}_{ab}$ are
found using
\bea
K^{(1),h}_{ab}&=& \varepsilon_{abcd} \varepsilon^{\mu\nu\rho\s}{(D^{(0)}_\mu
A_{\nu})^c}_ e\,A_\rho^{ed}\,e^{-1}{e_\s}^h,\\
K^{(i),h}_{ab}&=& \varepsilon_{abcd} \varepsilon^{\mu\nu\rho\s}
\w^{(i-1)cf}_\mu\, A_{\nu fe}\,A_\rho^{ed}\,e^{-1}{e_\s}^h,\nn\\ &&\quad (i=2,3,...)
\eea
where $e=\det({e_\mu}^a)$ and $D^{(0)}$ is the covariant derivative with
respect to the connection $\w^{(0)}_{ab}$.


\end{document}